\documentclass[pre,twocolumn,groupedaddress,floatfix]{revtex4-1}
\usepackage{amssymb,amsmath}
\usepackage{graphicx}
\usepackage{subfigure}
\usepackage[english]{babel} 
\usepackage{float}
\usepackage{color}

\usepackage{lipsum}
\begin{document}
\newcommand{\be}{\begin{equation}}
\newcommand{\ee}{\end{equation}}
\newcommand{\rojo}[1]{\textcolor{red}{#1}}

\title{The fractional nonlinear electrical lattice}

\author{Mario I. Molina}
\affiliation{Departamento de F\'{\i}sica, Facultad de Ciencias, Universidad de Chile, Casilla 653, Santiago, Chile}

\date{\today }

\begin{abstract} 
We examine the linear and nonlinear modes of a one-dimensional nonlinear electrical lattice, where the usual discrete Laplacian is replaced by a fractional discrete Laplacian. This induces a long-range intersite coupling that, at long distances,  decreases as a power law. In the linear regime, we compute both, the spectrum of plane waves and the mean square displacement (MSD) of an initially localized excitation, in closed form in terms of regularized hypergeometric functions and the fractional exponent. 
The MSD shows ballistic behavior at long times, MSD$\sim t^2$ for all fractional exponents. When the fractional exponent is decreased from its standard integer value, the bandwidth decreases and the density of states shows a tendency towards degeneracy. In the limit of a vanishing exponent, the system becomes completely degenerate. For the nonlinear regime, we compute numerically the low-lying nonlinear modes, as a function of the fractional exponent. A modulational stability computation shows that, as the fractional exponent decreases, the number of electrical discrete solitons generated also decreases, eventually collapsing into a single soliton.

\end{abstract}

\maketitle

{\em Introduction}. It's been quite a long time since the earlier correspondence between Leibnitz and L'Hopital took place, concerning possible generalizations of the concept of a derivate and whether it made sense to ask questions such as: what is the a half derivate of a function? The basic starting point was the calculation of $d^{\alpha} x^{k}/ dx^{\alpha}$, where $\alpha$ is a non-integer number. This means
\be 
{d^{n} x^k\over{d x^n}}= {\Gamma(k+1)\over{\Gamma(k-n+1)}} x^{k-n} \rightarrow {d^\alpha x^k\over{d x^\alpha}} = {\Gamma(k+1)\over{\Gamma(k-\alpha+1)}} x^{k-\alpha}.\label{eq1}
\ee
From Eq.(\ref{eq1}) the fractional derivative of an analytic function $f(x)=\sum_{k} a_{k} x^{k}$ can be computed by deriving term by term. However, this basic procedure is not exempt from ambiguities. For instance, $(d^\alpha/d x^{\alpha})\ 1=(d^\alpha x^{0}/d x^\alpha)=(1/\Gamma(1-\alpha)) x^{-\alpha}\neq 0$, according to Eq.(\ref{eq1}). However, one could have also taken $(d^{\alpha-1}/d x^{\alpha-1})(d/dx)\ 1=0$. The initial studies were followed later by rigurous work by several people including Euler, Laplace, Riemann, and Caputo, to name some and  promoted fractional calculus from a mathematical curiosity to a full-blown research field\cite{fractional1,fractional2,fractional3}. Several possible definitions for the fractional derivative have been obtained, each one with its own advantages and disadvantages. One of the most used definitions is the Riemann-Liouville form
\be
\left({d^{\alpha}\over{d x^{\alpha}}}\right) f(x) = {1\over{\Gamma(1-\alpha)}} {d\over{d x}} \int_{0}^{x} {f(s)\over{(x-s)^{\alpha}}} ds
\ee
another common form, is the Caputo formula
\be
\left({d^{\alpha}\over{d x^{\alpha}}}\right) f(x) = {1\over{\Gamma(1-\alpha)}} \int_{0}^{x} {f'(s)\over{(x-s)^{\alpha}}} ds
\ee
where, $0<\alpha<1$. This formalism that extends the usual integer calculus to a fractional one, with its definitions of a fractional integral and fractional 
derivative, has found application in several fields: fluid mechanics\cite{fluid2}, fractional kinetics and anomalous diffusion\cite{metzler,sokolov,zaslavsky}, strange kinetics\cite{shlesinger}, fractional quantum mechanics\cite{laskin,laskin2}, Levy processes in quantum mechanics\cite{levy}, plasmas\cite{plasmas}, electrical propagation in cardiac tissue\cite{cardiac}, biological invasions\cite{invasion}, and epidemics\cite{epidemics}.
\begin{figure}[t]
 \includegraphics[scale=0.14]{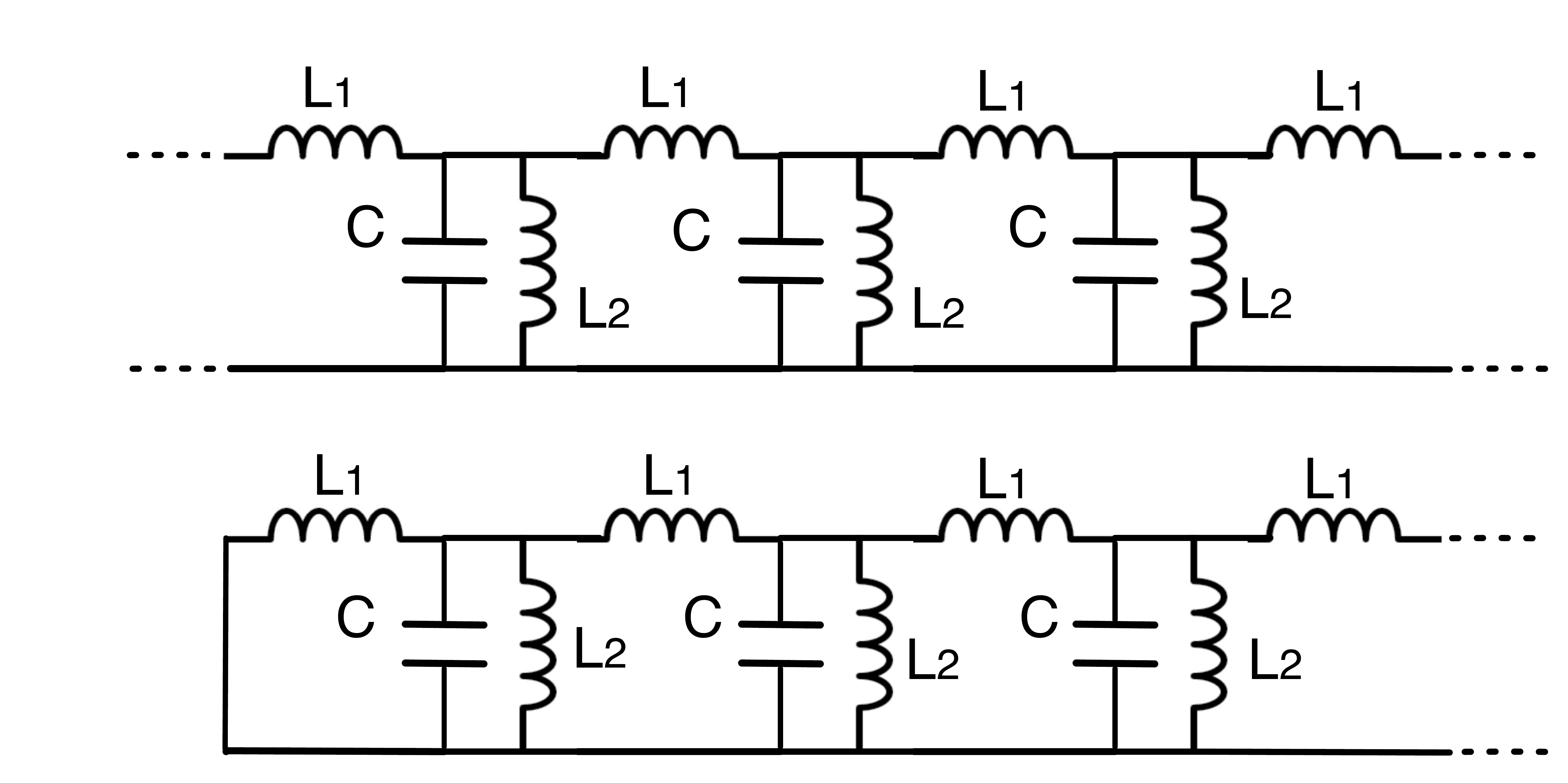}
  \caption{Infinite (top) and semi-infinite (bottom) 
  bi-inductive electrical lattice (after ref.\cite{circuit}).}  \label{fig1}
\end{figure}

On the other hand, one of the most interesting concepts in nonlinear physics is that of a soliton. It is a solitary wave solution of certain nonlinear differential  equations and is  characterized by having a spatial profile that remains undeformed upon time evolution. The origin of this behavior lies in a balance between dispersion and nonlinearity. Initially found as solutions of some system of coupled anharmonic oscillators, continuous and discrete solitons have by now been predicted and observed in a wide variety of settings: fluids\cite{fluidsA, fluidsB, turbulence}, biology (low frequency collective motion in proteins)\cite{davidov}, optics\cite{optical solitons1,optical solitons2,optical solitons3}, magnetism\cite{magnetism1,magnetism2}, nuclear physics\cite{nuclear}.  In particular, discrete solitons have also been predicted and observed in nonlinear electrical transmission lines\cite{electrical transmission lines1,electrical transmission lines2,lars1,lars2,lars3}. The reason is that a nonlinear electrical network can be regarded as a set of coupled anharmonic oscillators.

In this work, we aim at examining the consequences of the use of a fractional discrete Laplacian on the existence and stability of discrete soliton modes as well as on the transport of excitations in an electrical bi-inductive electrical network (Fig.1).  As we will see, fractionality changes the spectrum of plane waves, with a bandwidth that  decreases with decreasing $\alpha$ and a density of states who also decreases its width,  becoming completely degenerate when  $\alpha\rightarrow 0$. The discrete soliton phenomenology is more or less preserved, although the number of discrete solitons generated by the modulational stability mechanism depends strongly on the value of the fractional exponent. 

{\em The model}.\ Figure 1 shows a bi-inductive electrical lattice composed of a one-dimensional array of $LC$ circuits coupled inductively. $L_{1}$ and $L_{2}$ are the inductances and $C$ is a nonlinear capacitor, with a capacitance given by $C_{n}= C_{0} (1+\chi_{1}+\chi_{3} U_{n}^2)$ where $C_{0}$ is the capacitance in vacuum, $\chi_{1}$ is the linear susceptibility, $\chi_{3}$ is the third-order susceptibility,  and $U_{n}$ is the voltage drop. The nonlinear capacitance $C_{n}$ can be obtained by inserting a Kerr dielectric material between the capacitor plates. The electrical charge $Q_{n}$ on the $n$th capacitor is given by $Q_{n} = C_{n} U_{n}$. After using Kirchhoff's law, the equations for the voltages are
\be
{d^2 Q_{n}\over{d t^2}} = {1\over{L_{1}}}(U_{n+1}-2 U_{n}+U_{n-1}) - {1\over{L_{2}}}U_{n},
\ee
After inserting the expression for $Q_{n}$ in terms of $U_{n}$ and after introducing dimensionless variables, we obtain
\be
{d^2\over{d \xi^2}} \{\  (1+\chi_{1})\ V_{n} + \gamma \ V_{n}^3 \ \} = (V_{n+1}-2 V_{n}+V_{n-1})-\omega^2 \ V_{n},\label{eq5}
\ee
where $V_{n}=U_{n}/U_{c}$, $\gamma=\chi_{3} U_{c}^2$, $\omega^2=(\omega_{2}/\omega_{1})^2$, $\xi=\omega_{1} t$, where $\omega_{1}=1/\sqrt{L_{1} C_{0}}$, $\omega_{2}=1/\sqrt{L_{2} C_{0}}$ are the resonant frequencies and $U_{c}$ is a characteristic voltage.

The first term on the RHS of Eq.(\ref{eq5}), is the discrete Laplacian $\Delta_{n} V_{n}$ Thus, we can write
\be
{d^2\over{d \xi^2}} \{\  (1+\chi_{1})\ V_{n} + \gamma \ V_{n}^3 \ \} -\Delta_{n} V_{n}+\omega^2 \ V_{n}=0,\label{eq6}
\ee
We now promote this discrete one-dimensional Laplacian to its fractional form, by using results by Roncal et al.\cite{discrete laplacian} in which an expression is obtained for the $\alpha$-th power of the discrete Laplacian:
\be
(-\Delta_{n}^{\alpha}) V_{n} = \sum_{m\neq n} K^{\alpha}(n-m) (V_{n}-V_{m})
\ee
where
\be
K^{\alpha}(m) = L_{\alpha} {\Gamma(|m|-\alpha)\over{\Gamma(|m|+1+\alpha})},
\ee
with 
\be
L_{\alpha}={4^{\alpha} \Gamma(\alpha + (1/2))\over{\sqrt{\pi} |\Gamma(-\alpha)|}},
\ee
and $0<\alpha<1$ is the fractional exponent.

Thus, the main equation reads
\begin{eqnarray}
& & {d^2\over{d \xi^2}} \{\  (1+\chi_{1})\ V_{n} + \gamma \ V_{n}^3 \ \} + \sum_{m\neq n} K^{\alpha}(n-m) (V_{n}-V_{m})\nonumber\\
& &\hspace{1cm}+ \omega^2 \ V_{n}=0,\label{eq10}
\end{eqnarray}
As we can see, the immediate effect of a fractional discrete laplacian is to introduce nonlocal interactions via a symmetric kernel $K^{\alpha}(n-m)$. Using the relation $\Gamma(n+\alpha)=\Gamma(n) n^\alpha$, we obtain the asymptotic expression
$K^{\alpha}(m)\rightarrow 1/|m|^{1 + 2 \alpha}$, i.e., a power-law
decrease of the coupling with distance.

\noindent
{\em Stationary modes}. We look for the stationary modes, in the form $V_{n}(t) = V_{n} \cos(\Omega \xi + \phi)$. In order to keep things simple, we use the rotating-wave approximation (RWA), where in $\cos(\Omega \xi +\phi)^3$ we neglect the higher harmonic:
$V_{n}(t)^3=V_{n}^3 \cos(\Omega \xi +\phi)^3\approx (3/4) V_{n}^3 \cos(\Omega \xi +\phi)$.
The stationay equation becomes:
\begin{eqnarray}
& & -\Omega^2 \{\  (1+\chi_{1})\ V_{n} + (3/4)\gamma \ V_{n}^3 \ \} + \sum_{m\neq n} K^{\alpha}(n-m) (V_{n}-V_{m})\nonumber\\
& &\hspace{1cm}+ \omega^2 \ V_{n}=0.\label{eq11}
\end{eqnarray} 

Let us consider first the linear case ($\gamma=0$),
\be
 \Omega^2 \{\  (1+\chi_{1})\ V_{n} \ \} = \omega^2 V_{n}+ \sum_{m\neq n} K^{\alpha}(n-m) (V_{n}-V_{m}),
 \ee
and look for the dispersion relation of plane waves, $V_{n}=A\ e^{i k n}$. One obtains:
\be
\Omega^2 (1+\chi_{1}) = \omega^2 + 4 \sum_{q\neq 0} K_{q}^{\alpha} \sin^2(q k /2).
\ee
This expression can be recast in closed form as
\begin{eqnarray}
\lefteqn{(1+\chi_{1}) \Omega^2 = \omega^2 + 2 {\Gamma(2\alpha)\over{(\Gamma+1)\Gamma(\alpha)}}}\nonumber\\
&\times&  [ 1-\alpha ( e^{-i k} {_{2}}F_{1} (1,1-\alpha,\alpha+2; e^{-i k})+\nonumber\\
& & e^{i k} {_{2}}F_{1} (1,1-\alpha,\alpha+2; e^{i k}) )]\label{om2}
\end{eqnarray}
where the ${_{2}}F_{1}$ are the regularized hypergeometric functions. Figure 2 shows the
dispersion $\Omega^2(k)$ for several fractional exponents $\alpha$. At $\alpha=1$, the standard case, the band is contained between $\omega^2/(1+\chi_{1})$ and $(\omega^2+4)/(1+\chi_{1})$. As $\alpha$ is decreased towards $0$, the bandwidth decreases steadily and at $\alpha\rightarrow 0^{+}$, the band becomes contained between $\omega^2/(1+\chi_{1})$ and $(1+\omega^2)/(1+\chi_{1})$.
\begin{figure}[t]
 \includegraphics[scale=0.35]{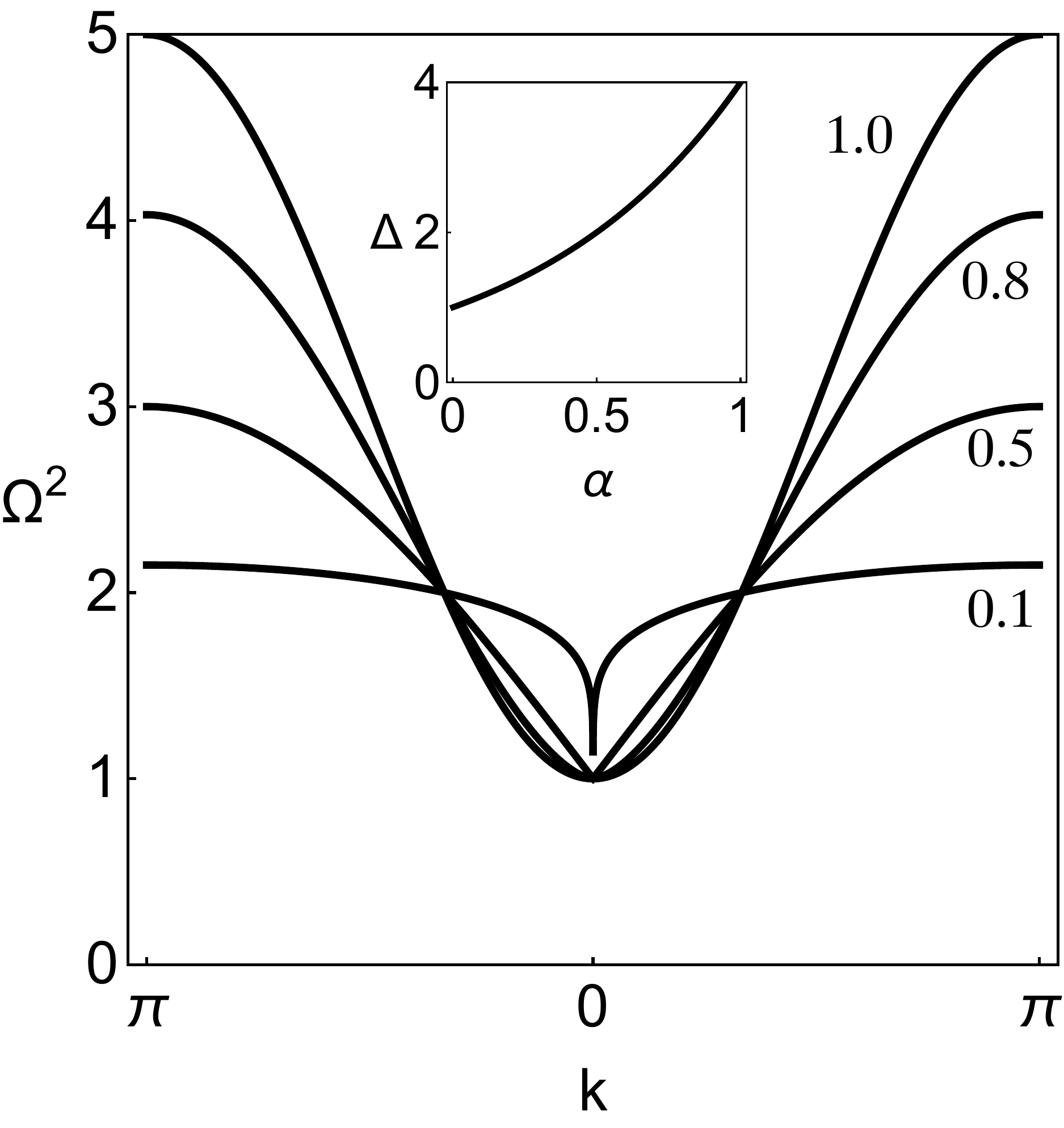}
  \caption{Dispersion relation for several fractional exponents, for $\chi_{1}=0$. 
  The number on each curve denotes the value of the fractional exponent $\alpha$. Inset: Bandwidth $\Delta$ as a function of fractional exponent, $\alpha$.}  \label{fig2}
\end{figure}
\begin{figure}[t]
 \includegraphics[scale=0.2]{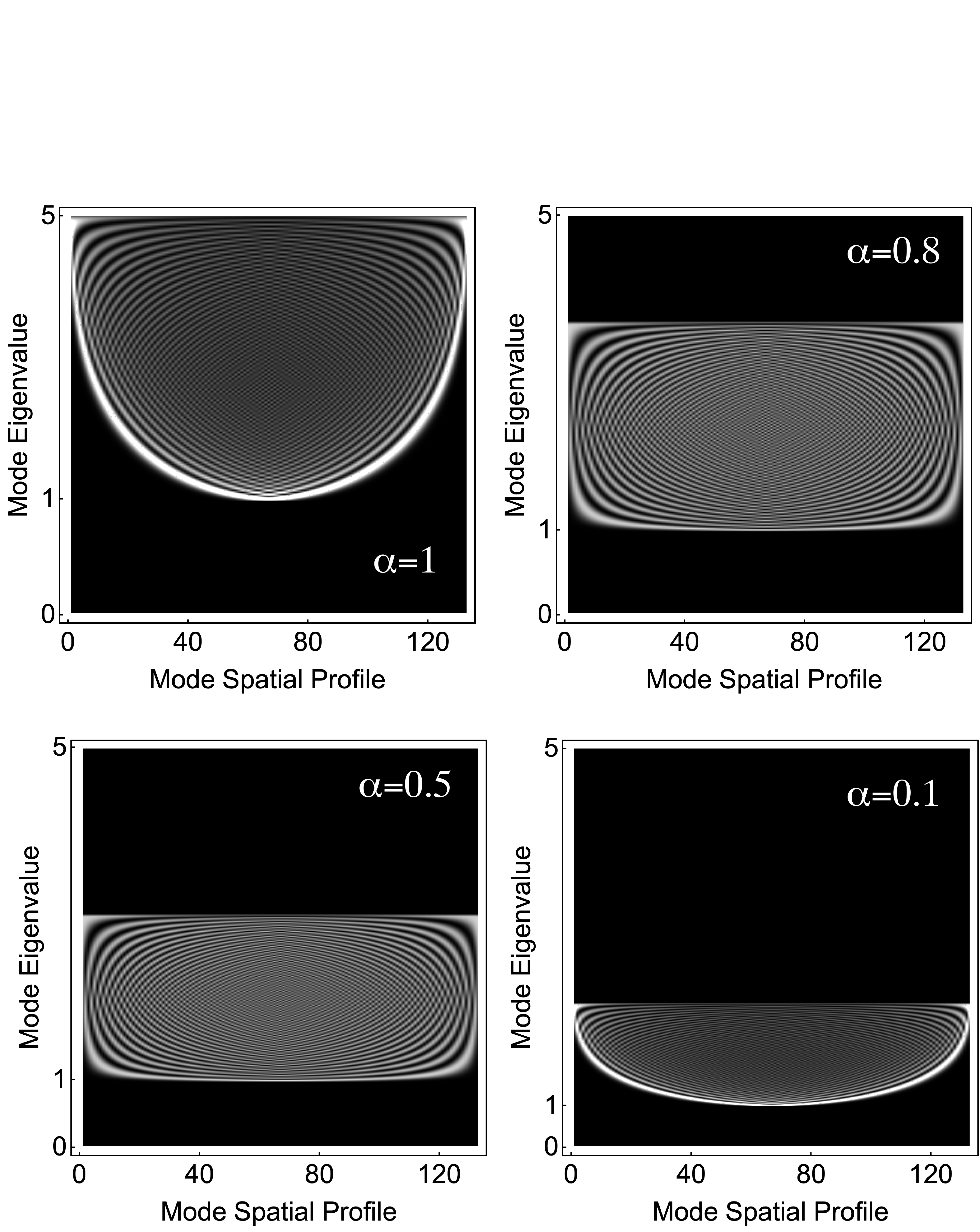}
  \caption{Density plot of the spatial profiles $|V_{n}|^2$ of the linear modes ordered according to their eigenvalue ($N=133, \omega=1, \chi_{1}=1$).}  \label{fig3}
\end{figure}
\begin{figure}[t]
 \includegraphics[scale=0.35]{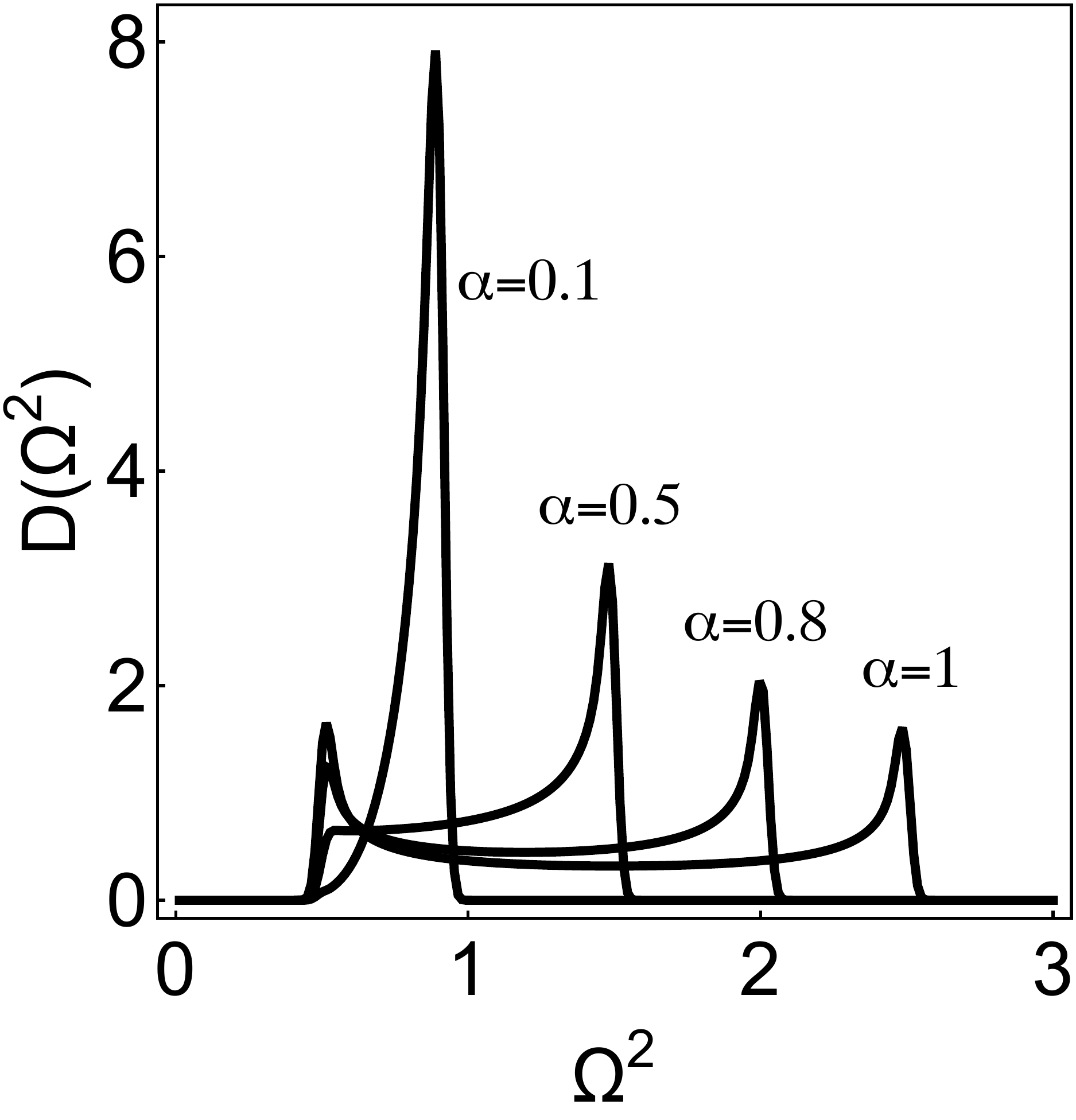}
  \caption{Density of states versus mode frequency for several fractional exponents. ($w=1, \chi_{1}=1, N=133$).}  \label{fig4}
\end{figure}
The bandwidth, defined by $\Delta(\alpha)=\Omega^{2}(\alpha,\pi)-\Omega^{2}(\alpha,0)$ is given in closed form by
\begin{eqnarray*}
\lefteqn{\Delta(\alpha)={4 \alpha \Gamma(1-\alpha)\Gamma(\alpha)\over{\pi (1+\chi_{1})}}\times}\\
& & \sin(\alpha \pi)\ [\  {{_2}F_{1}}(1,1-\alpha,2+\alpha,-1) +  {{_2}F_{1}}(1,1-\alpha,2+\alpha,1)\ ]
\end{eqnarray*}
and is shown in the inset of Fig.2.

Figure 3 shows the spatial profiles of all eigenmodes for several fractional exponents. 
Clearly, as $\alpha$ decreases the eigenvalues become more and more confined to an ever- decreasing energy range. In the limit $\alpha\rightarrow 0$, all the eigenvalues become confined to the band $\omega^2/(1+\chi_{1})<\Omega^2<(1+\omega^2)/(1+\chi_{1})$.
On the other hand, figure 4 shows the density of states (DOS)
\be
D(\Omega^2) = (1/N)\sum_{m} \delta(\Omega^2-\Omega_{m}^2)\label{density}
\ee
where $N$ is the number of sites and the sum is over all modes. We notice that, for all
fractional exponents, the DOS displays expected van Hove singularities, of finite height due to the finite-size effects. As the fractional exponent decreases from the standard  case ($\alpha=1$), the DOS high energy boundary starts receding towards the low energy boundary, which stays always in place during the process. Due to normalization, this narrowing process also increases the height of the DOS. In the limit of $\alpha\rightarrow 0$, the DOS diverges at 
$\Omega^2=\omega^2/(1+\chi_{1})$, and the system becomes completely degenerate.

Let us look at the mean square displacement (MSD), that serves to 
monitor the propagation of electrical excitations. The MSD is defined as
\be
\langle n^2 \rangle = \sum_{n} n^2 |V_{n}(\xi)|^2 / \sum_{n} |V_{n}(\xi)|^2
\label{MSD}
\ee
For a completely localized initial voltage $V_{n}(0)=A\ \delta_{n 0}$ and no currents, $(d V_{n}/d \xi)(0)=0$, we have formally
\begin{eqnarray*}
 V_{n}(\xi) &=& (A/4 \pi) \int_{-\pi}^{\pi} e^{i (k n-\Omega_{k})\xi} dk\\
            & & + (A/4 \pi) \int_{-\pi}^{\pi} e^{i (k n+\Omega_{k})\xi} dk
\end{eqnarray*}
where $\Omega_{k}$ is given by Eq.(\ref{om2}). After replacing this form for $V_{n}(\xi)$ into Eq.(\ref{MSD}), one obtains after some algebra, a closed form expression for $\langle n^2 \rangle$:
\be
\langle n^2 \rangle = {(1/2 \pi)\int_{-\pi}^{\pi}d k (d \Omega_{k}/d k)^2 (1 - \cos(2\ \Omega_{k}\ \xi))\ \xi^2 
\over{1 + (1/2\pi) \int_{-\pi}^{\pi} d k\ \cos(2\ \Omega_{k}\ \xi)}}.\label{n2closed}
\ee
As we can see from the structure of Eq.(\ref{n2closed}), as time $\xi$ increases, the contributions from the cosine terms to the integrals decrease and, at long times, $\langle n^2\rangle$ approaches a ballistic behavior
\be
\langle n^2 \rangle =\left[ {1\over{2 \pi}} \int_{-\pi}^{\pi} \left( {d \Omega(k)\over{d k}}\right)^2\ dk\right]\ \xi^2\hspace{1cm} (\xi\rightarrow \infty),\label{RMS}
\ee
while at short times,
\be
\langle n^2 \rangle = \left[ {1\over{2 \pi}} \int_{-\pi}^{\pi} \left( \Omega_{k}{d \Omega_{k}\over{d k}} \right)^2 dk\right] \ \xi^4 \hspace{1cm}(t\rightarrow 0).
\ee
Since the transport exponent is defined as the one corresponding to the dominant behavior at long times, we conclude that the asymptotic transport of our system is ballistic: $\langle n^2 \rangle \sim g(\alpha) \xi^2$, where we can identify $\sqrt{g(\alpha)}$ as a kind of characteristic `speed' for the ballistic propagation. This speed depends implicitly on the fractional exponent through $\Omega_{k}^2$, but the ballistic exponent is valid for all $0<\alpha<1$. 

{\em Nonlinear electrical modes}. We now turn our attention to the 
nonlinear modes of the system, which are solutions to Eq.(\ref{eq11}), with $\gamma\neq 0$. These equations constitute a system of nonlinear coupled difference equations, of the form ${\vec F} (\vec{V}) = 0$, 
where $\vec{V} = (V_{1},V_{2},\cdots,V_{N})$.
\begin{figure}[t]
 \includegraphics[scale=0.375]{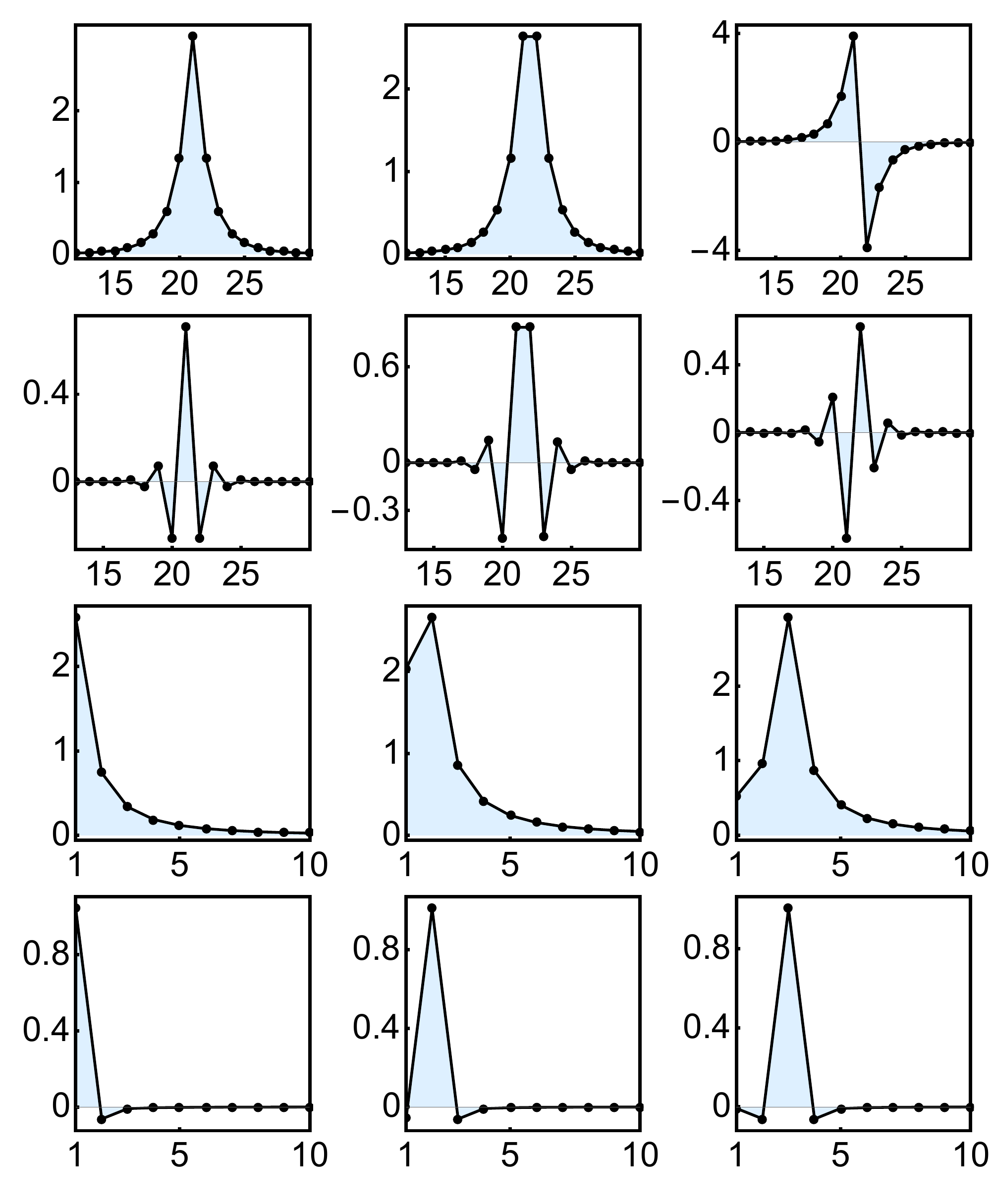}
  \caption{Examples of some low-lying nonlinear modes of the electrical lattice for $\chi_{1}=0$ and fractional exponent $\alpha=0.5$. On each plot, the vertical axes denotes the mode amplitude, while the horizontal axis indicate positions along the electrical array.
  First row: Bulk modes for $\chi_{3}=1$. (a), (b) and (c) denotes the `odd', `even' and `twisted' modes. Second row: Bulk modes for $\chi_{3}=-1$. (d), (e) and (f) denotes the `odd', `even' and `twisted' modes. Third row: Surface modes for $\chi_{3}=1$. (g), (h) and (i) denotes first layer, second layer and third layer modes. Fourth row:  Surface modes for $\chi_{3}=-1$. (j), (k) and (l) denotes first layer, second layer and third layer modes.}  \label{fig5}
\end{figure}
\begin{figure}[t]
 \includegraphics[scale=0.35]{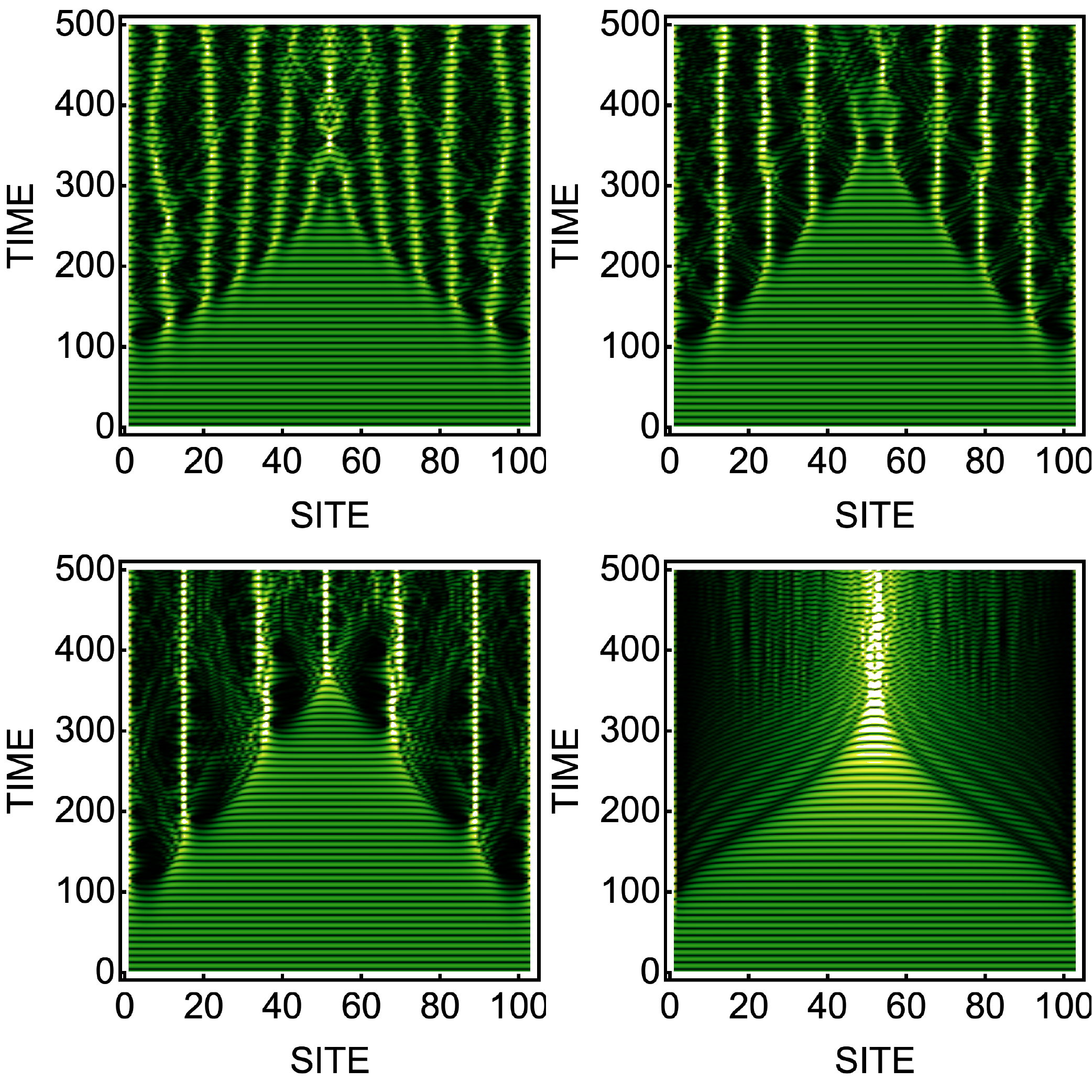}
  \caption{An example of modulational stability for the electrical array, for several fractional exponent values: Top left: $\alpha=1$. Top right: $\alpha=0.7$. Bottom left: 
  $\alpha=0.5$. Bottom right: $\alpha=0.1$. In all cases $\chi_{1}=0, \chi_{3}=1.$}  \label{fig6}
\end{figure}
Numerical solutions are obtained by using a multidimensional Newton-Raphson scheme. This method solves the system of nonlinear equations starting from a seed, which is supposed to be an approximate solution. If the initial guess is close to the real solution, convergence to the solution will be quickly reached. For instance, to find the fundamental solution in
the bulk which has a single localized maximum, we use as a guess something like $(0,0,\cdots,A,0,0,\cdots)$. For the
fundamental surface mode, we would use $(A,0,\cdots 0,0)$. Figure 5 shows examples of some low power bulk and surface modes obtained from this procedure. These soliton modes lie outside the band, i.e., in the gaps $0<\Omega^2<\omega^2/(1+\chi_{1})$ or $(\omega^2+4)/(1 + \chi_{1})<\Omega^2<\infty$. For a given exponent $\alpha$, the spatial profiles of all these modes are qualitatively similar to the ones found for the standard integer case ($\alpha=1$). 

Now, given a set of nonlinear electrical modes, we wonder about the linear stability of these modes. A particularly interesting case is the calculation of the modulational stability of a uniform array, where the initial condition consists of a constant voltage at all sites of the array, i.e., a flat front $V_{n}=A$. Such solution is always possible, provided $1+\chi_{1}+(3/4) \gamma A^3 >0$, for all $\alpha$, according to Eq.(\ref{eq11}). The idea is to start with the uniform state and compute whether after some time, the profile keeps its form. When it does not, the instability is termed a  modulation instability (MI). As an example, let us compute the MI for our electrical system. We use an array of $N=104$ units, a maximum evolution time of $\xi_{max}=500$, an initial amplitude of $A=1$, $\chi_{1}=0$ and a nonlinear susceptibility of $\chi_{3}=1$. The fractional exponent chosen is $\alpha=0.1$, a value substantially away from the standard case ($\alpha=1$). At both extremes of the array, we perturb the amplitudes as $A_{1}\rightarrow 1.01 A_{1}, A_{N}\rightarrow 1.01 A_{N}$.
Results are shown in Fig.6, where we show density plots where the height represents the amplitude of the voltage, while the vertical and horizontal axes denote the time and the position along the array, respectively. Each plot corresponds to a different fractional exponent. In all cases we observe the presence of modulational instability where the uniform front collapses after some time into a number of filament-like structures, that can be interpreted as discrete solitons. Indeed, previous works on the DNLS have identified  MI as a mechanism for the creation of discrete solitons. The same phenomenon seems to be at work here. We observe that as the fractional exponent decreases, the number of discrete solitons created decreases. At small $\alpha$, only one filament remains. The solitons generated are more or less equidistant from each other and no merging of them is observed for the times explored.
 
{\em Conclusions}\ We have examined the effect of using a fractional definition of the Laplacian for a one-dimensional array of coupled nonlinear electrical units. The introduction of fractionality gives rise to a nonlocal coupling between the units, which at long distances, decreases as a power law. 

In the linear regime, and in the presence of fractionality, there are modes in the form of electric plane waves whose dispersion relation was computed in closed form in terms of hypergeometric functions. The main effect of a fractional exponent is the reduction of the bandwidth with decreasing exponent. The density of states shows that, as the exponent decreases, the states shifts to the lower band edge. In the limit of a vanishing fractional exponent, all the states become completely degenerate, at an energy value proportional to the ratio of the fundamental resonant frequencies of the array. The mean square displacement of an initially localized electric excitation was calculated in closed form, showing a ballistic behavior at long times, while at short times the behavior was quartic in time. 

In the nonlinear regime, the nonlinear electric modes were computed  with the help of the rotating-wave approximation. Their bulk and surface profiles were similar to their standard, not fractional  counterpart. Finally, the modulational stability of the array was computed numerically. As expected, the onset of instability gave rise to a number of localized spatial structures that persisted in time, thus generating discrete solitons, as in the standard case.  These electric solitons form a sort of mini-array and its number depended strongly upon the value of the fractional exponent, decreasing their number as the exponent decreases. In the limit of a small exponent, only a single soliton remains in place.

We conjecture that these effects could be ascribed  to the particular form of the long-range coupling among the electrical units, induced by fractionality.

\acknowledgments
This work was supported by Fondecyt Grant 1200120.

\end{document}